# A Fault Location Method Based on Electromagnetic Transient Convolution Considering Frequency-Dependent Parameters and Lossy Ground

Guanbo Wang, *Student Member, IEEE,* Chijie Zhuang, *Member, IEEE*, Jun Deng, and Zhicheng Xie

*Abstract*—As the capacity of power systems grows, the need for quick and precise short-circuit fault location becomes increasingly vital for ensuring the safe and continuous supply of power. In this paper, we propose a fault location method that utilizes electromagnetic transient convolution (EMTC). We assess the performance of a naive EMTC implementation in multi-phase power lines by using frequency-dependent parameters in real fault simulation, while using constant parameters in pre-calculation. Our results show that the location error increases as the distance between the fault location and the measurement location increases. Therefore, we adopt the aerial mode transients after phase-mode transformation to perform the convolution, which reduces the influence of frequency-dependence and ground loss. We conduct numerical experiments in a 3-phase 100-km transmission line, a radial distribution network and IEEE 9-bus system under different fault conditions. Our results show that the proposed method achieves tolerable location errors and operates efficiently through direct convolution of the real fault-generated transient signals and the pre-stored calculated transient signals.

## I. INTRODUCTION

As the capacity of power systems increases, the need for quick and accurate short-circuit fault location becomes more important for safe and continuous power supply. Existing fault location methods can be mainly divided into two categories: impedance-based [1][2] and traveling wave-based [3]-[6]. Impedance-based methods are straightforward, but their accuracy can be easily affected by the fault impedance and network structure. Traveling wave-based methods have better accuracy, but they require high-frequency signal acquisition equipment, which can increase costs. In recent decades, a new type of traveling wave-based fault location method, known as Electromagnetic Time Reversal (EMTR) based methods, has been proposed and has gained extensive research interest [7]-[10]. Compared to traditional traveling wave-based methods, EMTR fault location methods make use of the full fault-generated transient instead of relying on arriving time instants. Because of this, EMTR methods are potentially more robust and accurate.

The idea for this paper originates from the EMTR-based method, which was first applied to transmission line fault location in 2012 [7]. Since then, numerous studies on the practicability of EMTR-based methods have been conducted through numerical simulations and real experiments [8]-[13]. A numerical test of a complex T-network composed of cables with different wave impedances was carried out in [8], which demonstrated that accurate short-circuit fault location can be achieved with a single-ended signal measurement. Moreover, [9] considered line loss using constant line parameters and showed that it has little influence on fault location accuracy for transmission lines less than 50 km in length. Additionally, experiments on the EMTR fault location method have been conducted. In [12] and [13], Wang et al. conducted fault location experiments in medium voltage distribution networks in China and Switzerland, respectively. These experiments demonstrated that fault locations can be correctly identified. However, these experiments were carried out in small-scale power networks, using lines less than several kilometers in length.

In general, the EMTR-based fault location process can be divided into two stages: the forward process and the reverse process. During the forward process, fault-generated transient signals are collected or measured at one or two ends of the transmission line. During the reverse process, a series of assumed short-circuit branches are set along the transmission line as the guess fault locations (GFLs). The transient signals are then injected back into the transmission line after time reversal. The fault current signal energies (FCSEs) are numerically calculated, and the maximum point of the FCSEs determines the real fault position.

The first motivation of this work is that in the reverse process, the FCSEs are calculated repeatedly for each guessed fault location, resulting in repeated calls to electromagnetic transient simulation software. In [14], a new fault location method using direct convolution of the fault-generated transients was proposed, which only requires the pre-calculation of assumed transients at the GFLs once for a given network, and can be used repeatedly for multiple faults. Derivation and numerical experiments of this method were carried out using single-phase transmission lines, while the effectiveness of the method in multi-phase lines needs validation.

We want to emphasize that in real applications, during the forward process, the fault-generated transient signals propagate through real power lines and are then measured. The power line parameters are frequency-dependent and may be influenced by the lossy ground. However, in the reverse process, the FCSEs are often calculated using constant or averaged parameters [7]-[13]. This is also the case in many existing literatures on fault location methods, such as [15]-[18]. An accurate wave velocity is crucial for fault location using traveling wave-based methods. However, in real applications, the wave propagation velocity (or transmission line parameters) is usually measured under a specific frequency. Therefore, it is not common to consider the frequency-dependent wave velocity, and constant parameters are often used. It's worth noting that in [12] and [13], the lines were relatively short and the ground resistivity was low. For instance, the transmission line in [12] was 677 meters long, and the ground resistivity was 10 Ω·m, while the fault distance in [13] was less than 3.6 km. Therefore, the effects were reduced, presenting only a slight influence on the evaluated techniques. While this may not be a significant issue for short lines, as shown in [8], [12], and [13], however, for long transmission lines, we can expect that the location error will increase due to the frequency dependence of the wave velocity, as we will demonstrate later.

The main contribution of the paper includes:
(1) We propose a fault location method based on electromagnetic transient convolution (EMTC) that only requires a convolution of the measured transient signal and a pre-computed signal, and further extend this method to multi-phase transmission line fault locations.
(2) We consider the influence of frequency-dependent line parameters and lossy ground on the location accuracy, especially for relatively long transmission lines in areas with high ground resistivity. To address this, we introduce phase-mode transformation to reduce the location error. By using the aerial mode transients, it is safe to use frequency-independent constant line parameters (wave velocity) in the pre-calculation.

The rest of the paper is organized as follows: In Section II, we briefly review the basic principle of the EMTR-based fault location method and propose the fault location method using EMTC. In Section III, we extend the EMTC method to multi-phase lines for different types of faults and consider the influence of frequency-dependent parameters and lossy ground. We show that when constant-parameter lines are used in the pre-calculation, the location error grows roughly proportional to the fault distance. To address this issue, we introduce phase-mode transformation to correct the location error. In Section IV, we present the application of our proposed method to a 300-km real transmission line, a radial distribution network and IEEE 9-bus system under different fault conditions. Finally, we draw some conclusions in Section V.

## II. ELECTROMAGNETIC TRANSIENT CONVOLUTION (EMTC)-BASED FAULT LOCATION METHOD: SINGLE PHASE LINE CASES

The principle of fault location using EMTC originates from the concept of EMTR, which has been discussed in a series of papers [7]-[13]. In this paper, we introduce the classical EMTR-based fault location method to illustrate the key principle used in our proposed method.

### A. Basic Principle of fault location method based on EMTR

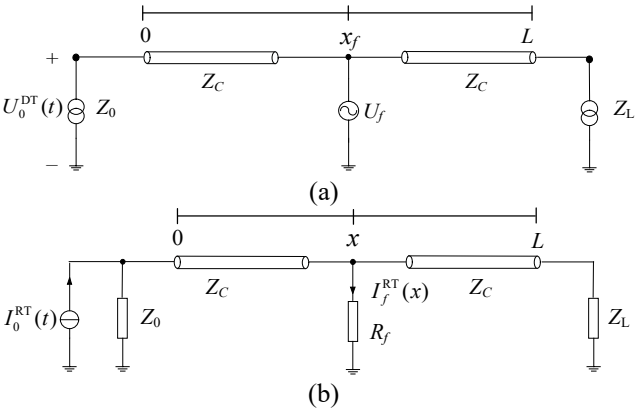

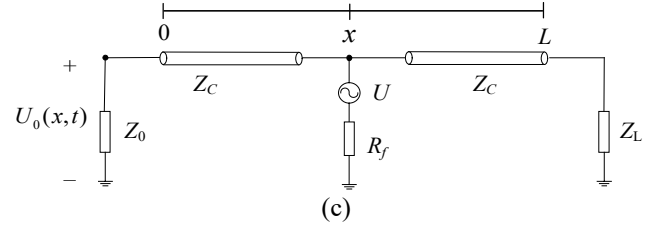

Fig. 1. (a) Representation of a fault along a single-conductor transmission line (forward process in EMTR), (b) the reverse process in EMTR, (c) fault location using EMTC.

Fig. 1(a) depicts the forward process (or in the *direct time*, DT) where a short-circuit fault occurs at $x = x_f$, represented by a voltage source $U_f$. $Z_0$ and $Z_L$ are the input impedances of the transformers. The frequency spectrum for fault-originated transients typically ranges up to hundreds of kilohertz. Hence, the input impedance behaves like a capacitance of several hundreds of pF [8] and is equivalent to a large impedance. We will analyze the effect of this simplification in Section IV. The high-frequency components of the transient can hardly pass through the transformers and interact with the source or load. Thus, the source impedance and load are not taken into consideration, and we only need to focus on the propagation and reflection between both ends. $Z_C$ represents the surge impedance of transmission lines. As in [18], the transient at one end of the line in the frequency domain $U_0^{DT}(\omega)$ can be expressed as:

$$U_0^{DT}(\omega) = \frac{(1+\rho_0)e^{-\gamma x_f}}{1+\rho_0 e^{-2\gamma x_f}} U_f(\omega), \quad (1)$$

where $\gamma = j\beta = j\omega/v$ for a lossless conductor, $v$ is the propagation speed and $\rho_0$ is the voltage reflection coefficient,

$$\rho_0 = \frac{Z_0 - Z_C}{Z_0 + Z_C}. \quad (2)$$

In the reverse process (or in the *reversed time*, RT), the signal is re-injected into the original system from the same end after the time-reversal operation and Norton equivalent, as shown in Fig. 1(b).

$$I_0^{RT}(\omega) = \frac{\left[U_0^{DT}(\omega)\right]^*}{Z_0}. \quad (3)$$

A series of priori assumed short-circuit branches are set along the line as the guessed fault locations (GFLs). Neglecting the fault impedance, the corresponding short-circuit current is

$$I_f^{RT}(x,\omega) = \frac{(1+\rho_0)^2 e^{-\gamma(x-x_f)}}{Z_0(1+\rho_0 e^{-2\gamma x})(1+\rho_0 e^{2\gamma x_f})} U_f^*(\omega). \quad (4)$$

The fault current signal energy (FCSE) reaches its maximum when $x_f' = x_f$. As a result, the fault can be located by calculating the fault current at different GFLs and determining the maximal FCSE.

### B. Fault location method using EMTC

From equation (4), we can observe that the maximum point of FCSE is independent of the energy-bounded signal $U_f(\omega)$

and the time reversal process. Therefore, if we opt to inject an arbitrary energy-bounded signal $U(\omega)$ (i.e., $0 < \int_{-\infty}^{+\infty} |U(\omega)|^2 d\omega < +\infty$) at the GFLs instead of calculating the time-reversed signal $I_0^{RT}(\omega)$ as depicted in Fig. 1(c), the transient voltage at the head end is

$$U_0(x,\omega) = \frac{(1+\rho_0)e^{-\gamma x}}{1+\rho_0 e^{-2\gamma x}} U(\omega). \quad (5)$$

The convolution of electromagnetic transients (1) and (5) is thus

$$C(x,\omega) = U_0(x,\omega) U_0^{DT}(\omega)$$
$$= \frac{(1+\rho_0)^2 e^{-\gamma(x_f+x)}}{(1+\rho_0 e^{-2\gamma x_f})(1+\rho_0 e^{-2\gamma x})} U_f(\omega) U(\omega). \quad (6)$$

Note that the magnitude of the EMTC (6) will have the same structure as (4), we can expect that the convoluted signal energy (CSE) also reaches its maximum at $x = x_f$ regardless of the choice of $U(\omega)$. This observation suggests that we can inject an arbitrary signal into the network in simulation and perform fault location by calculating an EMTC (By 'arbitrary', we mean that there is a vast range of signals available for the fault location process that are distinct from the time-reversed signal):

(1) [pre-calculation] For a given network, we can set a sequence of guessed short-circuit branches with an arbitrary excitation source along the line as the GFLs and then calculate the transient at one end of the line. We can store the results obtained in this manner, and this can be performed before an actual fault occurs.
(2) When an actual fault occurs, the transient signal generated by the fault is collected at the same end of the line as in (1).
(3) Calculate the EMTC of each pre-stored transient in (1) and the real transient signal collected in (2). Then, calculate the CSE of each EMTC, and the maximum CSE corresponds to the actual fault position.

To validate the EMTC algorithm, we carried out numerical experiments on a 20-km single-phase line. We used PSCAD/EMTDC with a time step of 0.1 μs to generate the transient, and MATLAB was used to implement the EMTC simulations. However, for real applications, EMTC can be implemented in embedded hardware. The parameters of the power line are shown in Table I, and we used 10-kΩ large resistances to represent the power transformers at both ends of the line.

TABLE. I. SINGLE-PHASE TRANSMISSION LINE PARAMETERS

| Parameter | Value |
|---|---|
| Length | 20 km |
| Wire diameter | 1 cm |
| Wire conductivity | 5.8*10$^7$S/m |
| Ground permittivity | 10 |
| Ground conductivity | 10$^{-1}$S/m |

In the pre-calculation, short-circuit branches with 1-Ω fault resistance and a lightning impulse

$$u(t) = 10\left(e^{-t/\alpha} - e^{-t/\beta}\right) kV, \alpha = 20\mu s, \beta = 3\mu s, \quad (7)$$

are set along the line at every GFLs. We then calculated the transient voltage at the same end of the line under the excitation of (7) for each short-circuit branch.

To simulate the fault, we set short-circuits with an impedance of 1 Ω along the line, respectively, and collected the fault-generated transient voltage signals at one end of the line for each short-circuit.

We want to emphasize that although we used the transmission line parameters in our simulations, what the EMTC algorithm requires is the propagation constant γ, or more precisely, an accurate wave velocity of the transient. This is also a critical requirement for other traveling wave-based methods. It can be understood from the telegraph equations that describe wave propagation, as well as equations (5)-(6).

Fig. 2 shows the CSEs for different real and guessed fault locations, with the CSEs normalized by dividing them by the maximum energy. The maximum points of the CSE always appear at the real fault position, as shown in Fig. 2, which confirms the correctness of the method.

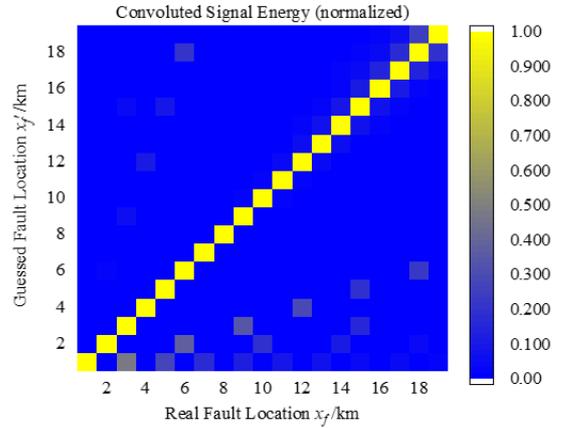

Fig. 2. The CSE at different GFLs for different faults. The fault was simulated using Bergeron models with a time step of 0.1 μs and the signal length used to do the EMTC was 5 ms.

While a rigorous theoretical proof of this method requires further investigation, we can consider why it works.

The transfer functions in equation (1) and (5) are

$$H^{DT}(\omega) = \frac{U_0^{DT}(\omega)}{U_f(\omega)} = \frac{(1+\rho_0)e^{-\gamma x_f}}{1+\rho_0 e^{-2\gamma x_f}}, \quad (8)$$

$$H(x,\omega) = \frac{U_0(x,\omega)}{U(\omega)} = \frac{(1+\rho_0)e^{-\gamma x}}{1+\rho_0 e^{-2\gamma x}}. \quad (9)$$

Wang et al. investigated the transfer functions (8) and (9) [19] and discovered the magnitude of $H^{DT}(\omega)$ reaches its local maximum when

$$f_k^{DT} = (2k+1)\frac{v}{4x_f}, k = 0,1,2,... \quad (10)$$

Similarly, $|H(x,\omega)|$ reaches its local maximum at $f_k$, where

$$f_k = (2k+1)\frac{v}{4x}, k = 0,1,2,... \quad (11)$$

When $x = x_f$, we have $f_k = f_k^{DT}, \forall k \in \mathbb{N}^*$, resulting in a maximum energy as all the local maxima of the two transfer functions are superposed. However, for other GFLs, the mismatch of local maxima causes the magnitude behavior to lose its harmonic characteristics. Hence, the effectiveness of EMTR fault location methods appears to be independent of $U(\omega)$, and is solely based on the magnitude characteristics of the transfer functions. Therefore, we can use a properly given signal in equation (5).

### III. EMTC Algorithm in Multi-Phase Lines Considering Frequency-Dependent Parameters and Lossy Ground

#### A. Naive implementation of EMTC in multi-phase lines

The derivation above is based on a single-phase line. Here, we extend the EMTC method to multi-phase cases, considering different fault types. The implementation details are presented in Table II.

TABLE. II. Naive Implementation of EMTC Algorithm for Multi-Phase Transmission Lines

| |
|---|
| **Input**: network parameters and topology; a given excitation signal $u(t)$ (e.g., a lightning impulse); a measured voltage signal $u_0^{DT}(t)$ generated by a fault |
| 1. Set a series of phase-to-ground, phase-to-phase and 3-phase short-circuit branches along the line as the GFLs, respectively. |
| 2. $u(t)$ is injected into the network at the GFLs. The voltage transient at one end of the line $u_0(x,t)$ from each GFL is calculated and stored. |
| 3. When the fault occurs, collect the fault-generated transient signal $u_0^{DT}(t)$ at the same terminal as in step 2. |
| 4. Recognize the fault type according to the waveform and amplitude of the transient. |
| 5. With the knowledge of fault type, calculate the EMTC of $u_0^{DT}(t)$ and the corresponding $u_0(x,t)$ of the same fault type, and the result is denoted as $c(x,t)$. |
| 6. Calculate the energy of $c(x,t)$ which is denoted by $E(x)$. |
| **Output**: the predicted fault location is $x_f = \arg|_x(\max(E(x)))$. |

Previous studies on EMTR-based fault location have used constant-parameter models in simulations to acquire electromagnetic transient signals in the reverse process, such as in [11]-[13]. However, in real-world applications, the parameters of actual transmission lines are frequency-dependent and can be affected by lossy ground. To illustrate the impact of this simplification, let us consider an example.

A 40-km, 3-phase overhead line is established in PSCAD/EMTDC using frequency-dependent models, as shown in Fig. 3. 10-kV sources are applied at the left end of the line, and 10-kΩ resistances are used to represent the power transformers at both ends of the line in the transient simulations. Phase-A-to-ground faults are assumed at $x = x_f$ along the line and the naive EMTC method is used to locate the faults. Note that the pre-calculation process is still carried out using constant-parameter models.

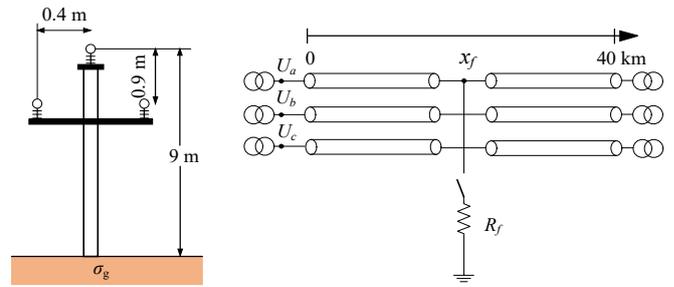

Fig. 3. The schematic representation for a 3-phase transmission line and a phase-to-ground fault. $\sigma_g$ is the ground conductivity.

To evaluate the impact of different ground conductivities on the EMTC method, we conducted simulations using the naive EMTC method, and the corresponding CSEs are shown in Fig. 4. The distance between two nearest GFLs depends on the expected location resolution and 10 m is adopted in this paper. In addition, all faults are created at one GFL, but this is not essential. If the fault is created between two GFLs, one of the two nearest GFLs will be identified as the fault location.

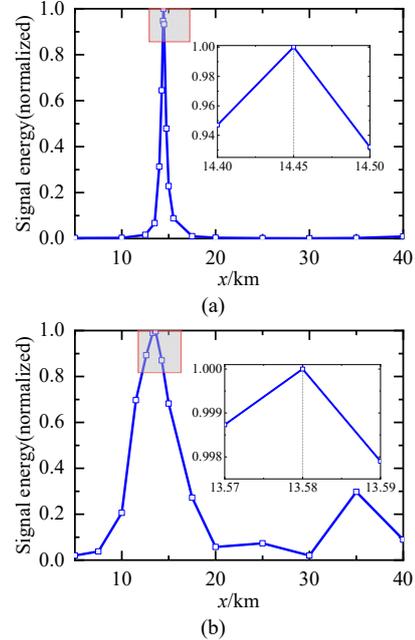

Fig. 4. CSE for fault at $x$=15 km, (a) $\rho_g = 10\ \Omega \cdot m$, (b) $\rho_g = 1000\ \Omega \cdot m$, line length is 40 km.

Fig. 4 indicates that the mismatch between transmission line models used in real fault simulation and pre-calculation can have an impact on the location accuracy, especially when the soil resistivity is high. For instance, in the case of phase-to-ground faults, an error of nearly 9% may occur when the ground resistivity is 1000 Ω·m.

For lossy transmission lines, transient signal energy (SE) dissipates as it propagates and tends to grow as the distance increases. As a result, the SE at $x=x_f$ (the actual fault position) decreases more than those at $x<x_f$, which may cause the location result to be slightly ahead of the real fault position.

Furthermore, Fig. 5 demonstrates that the location error is approximately proportional to the fault position $x_f$. As the ground resistivity increases, the location error also grows. However, the rough proportion relationship remains satisfied.

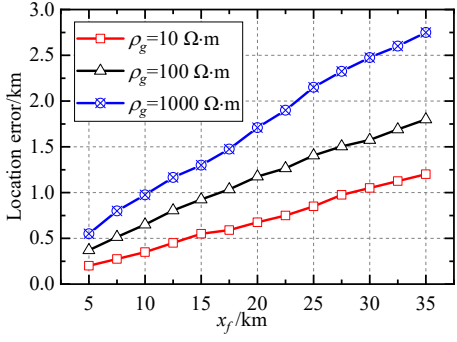

Fig. 5. The location error as a function of real fault position with different ground resistivities using naive implementation of EMTC.

### B. Discussion on the failure of naive EMTC implementation

Analysis of Fig. 5 suggests that location error is directly proportional to fault distance. Specifically, the location error can be expressed as a linear function of the actual fault position and the preliminary location result acquired using naive EMTC:

$$\Delta x_f = \lambda(x_f - x), \quad (12)$$

where $\Delta x_f$ represents the location error, $x_f$ represents the actual fault position and $x$ represents the preliminary location result obtained using naive EMTC.

Let's now take a brief look at the cause of the location error. Equations (10) and (11) show that when $x=x_f$, all local maxima of the two transfer functions are superimposed. However, when considering frequency-dependent parameters, the velocity becomes frequency-dependent. To see an example of this, Fig. 6 shows the frequency-dependent velocity of a single-conductor line with a height of 10 m and wire diameter of 1 cm. The corresponding expressions can be found in [20].

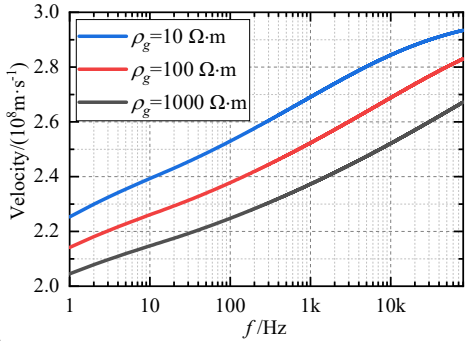

Fig. 6. Wave velocity for different ground resistivities. The wire conductivity is 5.8*10^7 S/m and the ground relative permittivity equals to 10.

The relationship shown in Fig. 6 is approximately logarithmic. Specifically,

$$v(f) = v_c \ln \frac{f}{f_0} + v(f_0). \quad (13)$$

where $v_c$ is a fitting coefficient.

The magnitude of the transfer functions (8) for frequency-dependent and frequency-independent parameters reaches local maximums at a series of frequencies $f_k^{DT,FD}$ and $f_k^{DT,FI}$, respectively,

$$f_k^{DT,FD} = (2k+1)\frac{v(f_k^{DT,FD})}{4x_f}, k=0,1,2,...$$

$$f_k^{DT,FI} = (2k+1)\frac{v^{FI}}{4x_f}, k=0,1,2,.. \quad (14)$$

where FD and FI represent the abbreviations of frequency-dependent and frequency-independent respectively, and $v^{FI}$ represent the velocity for lines with constant parameters.

Note that $f_k^{DT,FD} \approx f_k^{DT,FI}$, by Taylor's expansion,

$$v(f_k^{DT,FD}) \approx v(f_k^{DT,FI}) + \frac{v_c}{f_k^{DT,FI}}(f_k^{DT,FD} - f_k^{DT,FI}). \quad (15)$$

For each $k$, let $f_k = f_k^{DT,FD}$, we will have a series of preliminary location results $x_k$ under each given frequency $f_k^{DT,FD}$, and the corresponding location error $\Delta x_k$ can be estimated using (13)-(15),

$$\Delta x_k = \frac{v(f_k^{DT,FI}) - v^{FI}}{v(f_k^{DT,FI}) - v_c} x_f, k=0,1,2,..., \quad (16)$$

Table III shows that over 95% of the energy is concentrated within a frequency spectrum $[0, 20 f_0^{DT,FD}]$. As a result, we only need to consider values of $k$ less than 10 among all $k$.

Fig. 7 shows a comparison between different $\Delta x_k$ calculated using equation (16) and the real location error $\Delta x$ acquired through the naive EMTC method. The results indicate that $\Delta x$ is very similar to $\Delta x_0$, suggesting that the location error grows roughly linearly with distance.

TABLE. III. ENERGY RATIO OF $H^{DT}(\omega)$ WITH DIFFERENT FREQUENCY RANGE

| $x_f$/km | $f / f_0^{DT,FD}$ | | | |
| --- | --- | --- | --- | --- |
| | 10 | 15 | 20 | 25 |
| 10 | 0.852 | 0.919 | 0.955 | 0.975 |
| 20 | 0.856 | 0.922 | 0.957 | 0.976 |
| 50 | 0.858 | 0.924 | 0.958 | 0.977 |
| 100 | 0.858 | 0.924 | 0.959 | 0.978 |
| 500 | 0.854 | 0.920 | 0.957 | 0.977 |

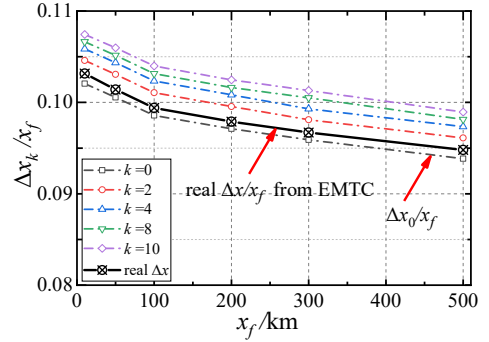

Fig. 7. The preliminary location errors $\Delta x_k$ in equation (16). The black solid line represents the real error $\Delta x/x_f$ acquired from naive EMTC method.

### C. EMTC in multi-phase lines with phase-mode transformation

Based on the derivation and analysis above, it is clear that the phase currents returning through the ground make the wave velocity frequency-dependent and can result in location errors, particularly when ground loss is included. To minimize the impact of a lossy ground, we utilize the aerial modes obtained after phase-mode transformation, which effectively reduces the influence of a lossy ground and improves the accuracy of fault location.

To illustrate the fault location process, we use a three-phase line as an example and consider three different fault types: phase-to-ground (P-G), phase-to-phase (P-P), and 3-phase (3P). To account for these faults, we pre-calculate the transients generated at the GFLs by the excitation source (e.g., (7)) and use Clarke's transformation [29] to separate the aerial and ground modes of the transient signal:

$$\begin{bmatrix} U_0 \\ U_\alpha \\ U_\beta \end{bmatrix} = T^{-1} \begin{bmatrix} U_a \\ U_b \\ U_c \end{bmatrix} = \frac{1}{3} \begin{bmatrix} 1 & 1 & 1 \\ 2 & -1 & -1 \\ 0 & \sqrt{3} & -\sqrt{3} \end{bmatrix} \begin{bmatrix} U_a \\ U_b \\ U_c \end{bmatrix}. \quad (17)$$

After performing the phase-mode transformation, the aerial mode transients are stored for later use. When a real fault occurs, the fault type is first identified by analyzing the fault-generated transients at the end. The EMTC is then carried out using the aerial modes of the fault-generated transients and the corresponding pre-stored signals to accurately locate the fault. The position of the actual fault is determined by the maximal CSE.

The flowchart of EMTC with phase-mode transformation is shown in Fig. 8.

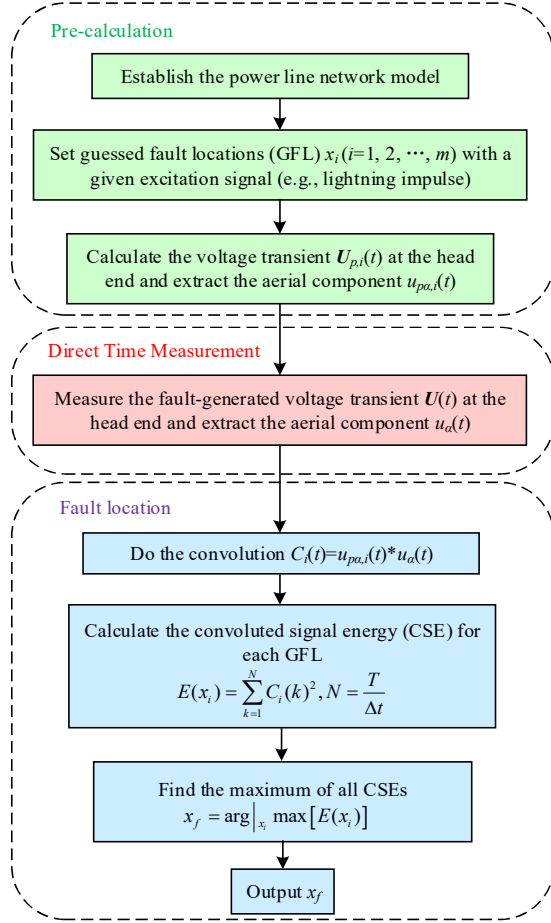

Fig. 8. The flowchart of improved EMTC-based fault location method.

To further test the effectiveness of our approach, we conducted numerical experiments using the setup shown in Fig. 3. Specifically, we set Phase-A-to-ground faults occurring at an angle of 90° along the transmission line and extended the length of the line to 100 km, with a fault impedance of 10 Ω. The lightning impulse (7) used in section II served as the excitation source for pre-calculation. To determine the CSEs, we utilized the EMTC of the pre-stored transients in conjunction with the real fault-generated transient voltage on aerial alpha mode.

For a P-G fault at 40 km, some specific transient signals required for fault location are shown in Fig. 9.

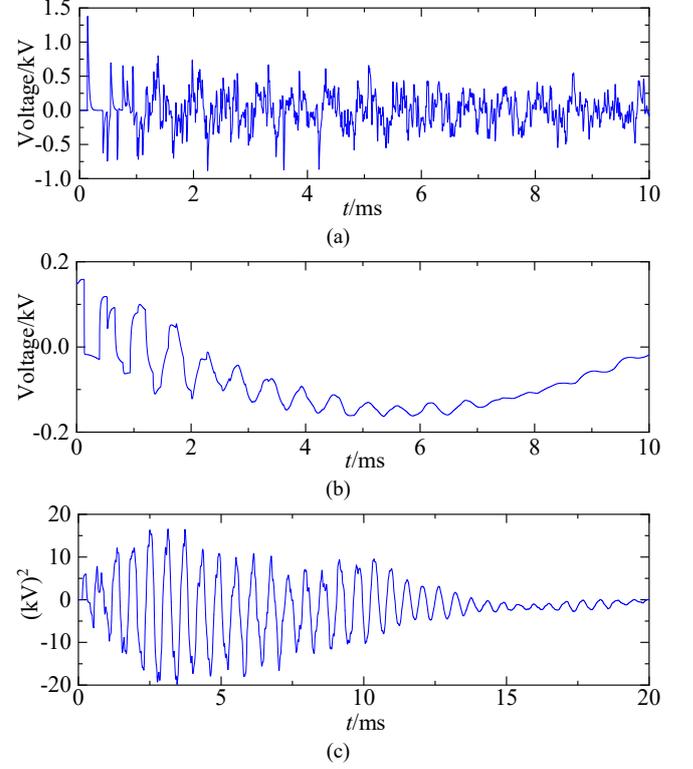

Fig. 9. The (a) pre-calculated aerial mode voltage signal, assuming a lightning impulse is injected at 40 km, (b) measured aerial mode voltage signal under P-G fault at 40 km, (c) corresponding convoluted signal of (a) and (b).

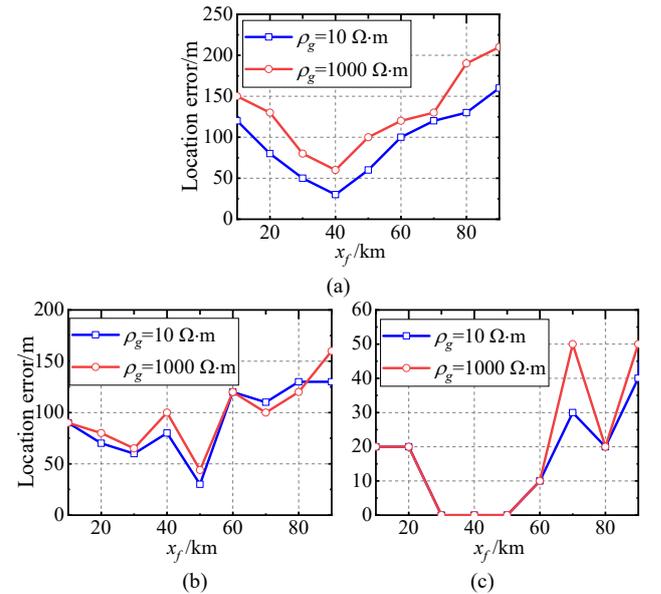

Fig. 10. Location error of (a) Phase-A-to-ground fault, (b) Phase-B-to-Phase-C fault, (c) 3-Phase fault.

The location errors for all P-G faults are presented in Fig. 10(a), it is evident that the location error for phase-to-ground faults is less than 0.2%, which represents a significant reduction compared to the error obtained in section III-A (9%). This reduction can be attributed to the negligible frequency dependence of the aerial mode propagation speed within the typical frequency spectrum of fault-generated transients. Consequently, the location errors are greatly reduced.

We also tested Phase-B-to-Phase-C faults along the line, with each fault separated by a distance of 10 km. The fault angle and impedance were set to 90° and 1 Ω, respectively, and the resulting location errors are presented in Fig. 10(b).

We also carried out tests on 3-phase faults, which were set along the line with each fault separated by a distance of 10 km. The fault impedance between each phase was assumed to be 1 Ω, and the fault angle and excitation source were set to 90° and (7), respectively. The resulting location errors are shown in Fig. 10(c) and are the least among the three fault types. This can be attributed to the fact that ground mode is more susceptible to frequency and ground loss. P-G and P-P faults have asymmetrical conductivity matrices, which means that the ground and aerial modes cannot be fully decoupled during phase-mode transformation. This results in a mixture of ground mode and aerial mode transients at the fault point, which can introduce location errors. However, in 3-phase faults, the influence of ground modes is almost entirely eliminated since they are symmetrical.

We wish to emphasize that EMTC algorithm is designed for maintenance, and the location result can be output after the operation of CBs. As a reference, the time cost of the algorithm carried out by MATLAB to determine the fault location in Fig. 10 for 10000 GFLs (the resolution is 10 m for a 100-km line) is about 20 s.

## IV. APPLICATION EXAMPLES

### A. Validation on a 220-kV transmission line

To validate our approach, we conducted numerical experiments on a 220-kV, 300-km 3-phase ideally-transposed transmission line with ground wires. The tower structure and line parameters are presented in Fig. 11 [21] and Table IV, respectively.

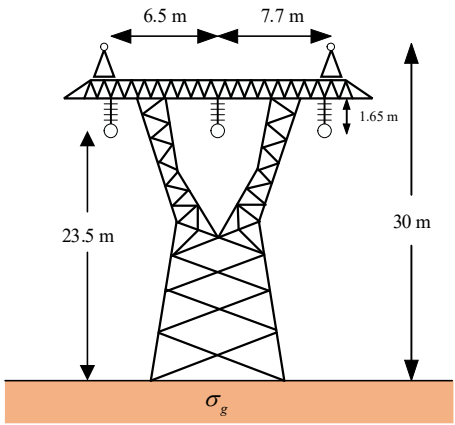

Fig. 11. Tower structure for a 220-kV 3-phase transmission line with 2 ground wires. $\sigma_g$ is the ground conductivity.

TABLE. IV. LINE PARAMETERS

| Parameter | Value |
|---|---|
| Conductor | 2*LGJ-400/35 |
| Ground wire | JLB4-150 |
| Sag | 11 m |

#### 1) Influence of transformer model

The previous derivation simplified the power transformer into a large impedance, neglecting the influence of the transformer on fault location accuracy. To estimate this influence, we introduced a high-frequency model inferred from typical 220-kV transformers [22] in the fault simulation, while both a realistic model and the simplified model were used in the pre-calculation process.

A phase-A-to-ground fault of 10 Ω was introduced at $x$=40 km, with a fault angle of 90° and a ground resistivity of 1000 Ω·m. Using EMTC, we obtained CSEs as shown in Fig. 12. Our results indicate that the impedance model provides sufficient accuracy for fault location, suggesting that the proposed simplified model is reasonable.

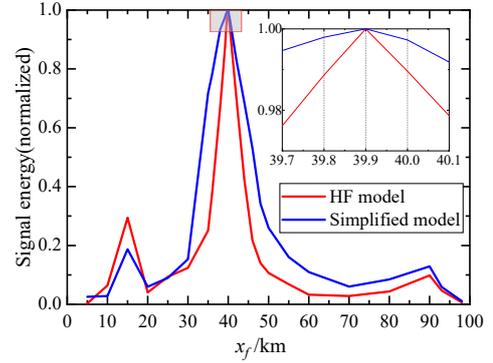

Fig. 12. CSE of a P-G fault at $x$=40 km. The CSEs from 100-300 km are concealed for visualization. The HF model is adopted from [22] and simplified model is a large impedance of 10 kΩ.

#### 2) Influence of transposition scheme

The proposed method utilizes aerial mode transients, therefore may be affected by transposition schemes. To estimate the influence, a series of P-G (most asymmetrical) faults are tested under ideally-transposed and actual transposed (three-line segments scheme) respectively. The location errors are shown in Table V. The fault angle and fault impedance are 90° and 100 Ω.

TABLE. V. FAULT LOCATION ERROR FOR DIFFERENT TRANSPOSITION SCHEMES (FAULT ANGLE AND IMPEDANCE ARE

| $x_f$/km | Ideal transposition | Actual transposition |
|---|---|---|
| 30 | 210 | 250 |
| 50 | 200 | 230 |
| 70 | 140 | 140 |
| 100 | 180 | 190 |
| 130 | 230 | 250 |
| 150 | 290 | 330 |
| 170 | 320 | 320 |
| 200 | 390 | 400 |
| 230 | 450 | 450 |
| 250 | 500 | 490 |
| 270 | 370 | 350 |

As is shown in Table V, the difference of location errors is within 10-m to 40-m. Modal transformation works effectively enough to decouple aerial and ground modes under both transposition schemes. Therefore, ideal transposition is adopted in the following for simplicity.

*3) Influence of different fault conditions*

The accuracy of fault location on transmission lines is affected by three primary factors, including fault impedances, fault angle, and fault position. These factors influence the wave reflection coefficient [23], [24], the amplitude of the fault-generated transient, and the propagation time, respectively. To evaluate the influence of these variables, we introduced different types of faults and located them using the EMTC method. We assumed a constant ground resistivity of 1000 Ω·m throughout the simulations. The following cases are considered:

(1) In general, the fault impedances for phase-to-phase (P-P) and three-phase (3P) faults may not be large. Therefore, to estimate the influence of fault impedances on fault location accuracy, we focused on phase-to-ground (P-G) faults, and we tested fault impedances of 1, 10, and 100 Ω.
(2) Different fault angles that are assumed to be 5° and 90° while other conditions remain the same.
(3) Different fault positions that vary from 30 km to 290 km for all 3 fault types.

The location errors of different fault conditions are summarized in Table VI. Our results demonstrate that the EMTC method is relatively insensitive to fault angles. Furthermore, our findings indicate that the location error is slightly higher for faults with higher impedances, due to the combined effects of waveform decay and deformation.

*4) EMTC performance compared with other metrics*

To provide a basis for comparison, we also evaluated the location results and error of a travelling wave-based fault location method using Time Difference of Arrival (TDOA). The classical two-end metric (18) and the setting-free metric (19) [25] were used in our evaluation, and the respective time instants are shown in Fig. 13.

$$x_{f,class} = 0.5 \cdot (L - v \cdot (t_L - t_0)), \quad (18)$$

$$x_{f,free} = \frac{t_{0r} - t_0}{(t_{0r} - t_0) + (t_{Lr} - t_L)} L . \quad (19)$$

To ensure accurate detection of time instants in both the classical two-end metric (18) and the setting-free metric (19) [25], we filtered the signals using differentiator-smoother filters [26]. The filtered signals were then compared with a threshold of 1% of the RMS voltage magnitude at the observation point [25]. Detection of the fault-generated traveling wave was triggered when the absolute value of the filtered signal exceeded the threshold. We used a sampling rate of 1 MS/s, identical to the EMTC method, to obtain the results shown in Fig. 14.

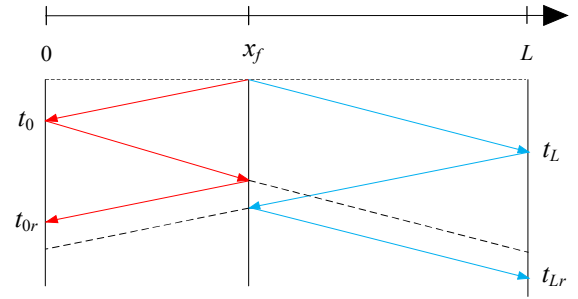

Fig. 13. Lattice diagram of fault-generated traveling waves measured for TDOA at the two terminals.

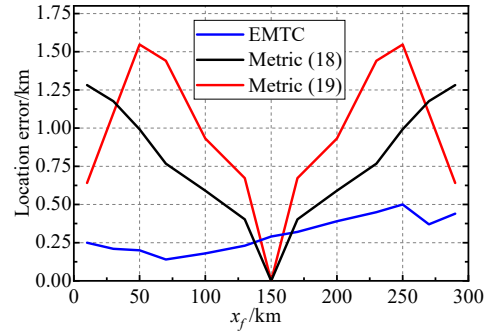

(a)

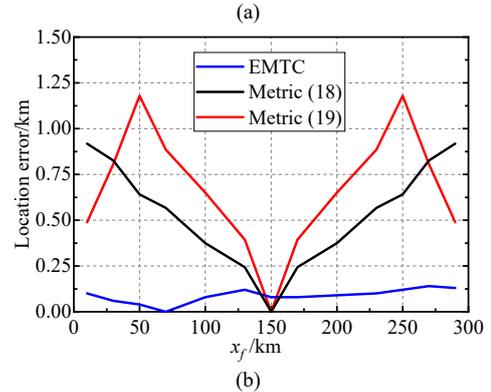

(b)

Fig. 14. The location error for (a) 90°/100-Ω P-G fault, (b) 5°/1-Ω P-P fault at different fault positions using different metrics. Line length is 300 km.

As demonstrated in Fig. 14, the location errors of the P-G (90°/100-Ω) and P-P (5°/1-Ω) faults, which are the worst cases of the EMTC method, are less than 2‰ and outperform the TDOA methods in average. For the faults at the middle of the line, EMTC works worse than metrics (18,19), which is expectable for a single-end method; however, the errors are still less than 1‰. Furthermore, the EMTC method offers the advantage of higher accuracy as a single-ended metric, without requiring any time synchronization measures. This is in contrast to TDOA-based methods, which rely on time synchronization and can be affected by synchronization errors.

TABLE. VI. FAULT LOCATION ERROR FOR DIFFERENT FAULT TYPES, FAULT POSITIONS, INCEPTION ANGLES AND FAULT IMPEDANCE

| $x_f$/km | Location error in meters | | | | | | | |
|---|---|---|---|---|---|---|---|---|
| | P-G/ 5°/10 Ω | P-G/ 90°/10 Ω | P-G/ 90°/1 Ω | P-G/ 90°/100 Ω | P-P/ 5°/1 Ω | P-P/ 90°/1 Ω | 3P/ 5°/1 Ω | 3P/ 90°/1 Ω |
| 30 | 140(0.47‰) | 140(0.47‰) | 140(0.47‰) | 210(0.7‰) | 60(0.2‰) | 60(0.2‰) | 40(0.13‰) | 40(0.13‰) |
| 50 | 80(0.27‰) | 80(0.27‰) | 80(0.27‰) | 200(0.67‰) | 40(0.13‰) | 40(0.13‰) | 20(0.07‰) | 20(0.07‰) |
| 70 | 50(0.17‰) | 50(0.17‰) | 20(0.07‰) | 140(0.47‰) | 0(0) | 0(0) | 0(0) | 0(0) |
| 100 | 120(0.4‰) | 120(0.4‰) | 120(0.4‰) | 180(0.6‰) | 80(0.27‰) | 80(0.27‰) | 0(0) | 0(0) |
| 130 | 120(0.4‰) | 120(0.4‰) | 120(0.4‰) | 230(0.77‰) | 120(0.4‰) | 120(0.4‰) | 0(0) | 0(0) |
| 150 | 100(0.33‰) | 100(0.33‰) | 100(0.33‰) | 290(0.97‰) | 80(0.27‰) | 80(0.27‰) | 0(0) | 0(0) |
| 170 | 90(0.3‰) | 90(0.3‰) | 80(0.27‰) | 320(1.07‰) | 80(0.27‰) | 80(0.27‰) | 0(0) | 0(0) |
| 200 | 140(0.47‰) | 140(0.47‰) | 140(0.47‰) | 390(1.3‰) | 90(0.3‰) | 90(0.3‰) | 10(0.03‰) | 10(0.03‰) |
| 230 | 170(0.56‰) | 170(0.56‰) | 170(0.56‰) | 450(1.5‰) | 100(0.33‰) | 100(0.33‰) | 10(0.03‰) | 10(0.03‰) |
| 250 | 230(0.77‰) | 230(0.77‰) | 200(0.66‰) | 500(1.67‰) | 120(0.4‰) | 120(0.4‰) | 20(0.07‰) | 20(0.07‰) |
| 270 | 330(1.1‰) | 330(1.1‰) | 230(0.77‰) | 370(1.23‰) | 140(0.47‰) | 140(0.47‰) | 20(0.07‰) | 20(0.07‰) |

## B. A radial distribution network

The second case study involves a radial distribution network, as illustrated in Fig. 15. The distribution network structure features multiple branches, resulting in more complex traveling wave attenuation and reflection, making it a worthwhile subject of analysis. The overhead line parameters for this study were the same as those used in Fig. 3. To model the transformer, we adopted the approach presented in [27].

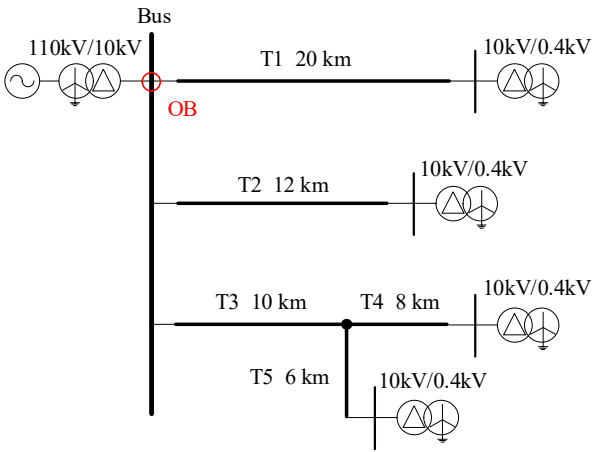

Fig. 15. A radial distribution network structure. The bus voltage on the low-voltage side of the transformer (OB) is monitored.

As previously discussed, the ground mode has the greatest impact on P-G faults. To evaluate the performance of the EMTC method under different P-G fault conditions, we assumed various P-G faults within the network and located them using the EMTC method. We held the fault angle and impedance constant, at 90° and 10 Ω, respectively. The bus voltage on the low-voltage side of the transformer is monitored.

To simulate the fault, we employed frequency-dependent models, while pre-calculation was carried out using constant-parameter models. Through this approach, we were able to estimate the location errors for different fault scenarios, as summarized in Table VI.

TABLE. VII. FAULT LOCATION RESULTS AND ERRORS USING EMTC

| Faulty line | Distance/km | Located faulty line | Location result/km | Error/m |
|---|---|---|---|---|
| T1 | 5 | T1 | 4.98 | 20 |
| T1 | 15 | T1 | 14.88 | 120 |
| T2 | 3 | T2 | 3.00 | 0 |
| T2 | 10 | T2 | 9.93 | 70 |
| T3 | 4 | T3 | 3.96 | 40 |
| T4 | 2 | T4 | 1.99 | 10 |
| T5 | 5 | T5 | 4.90 | 100 |

The results presented in Table VII demonstrate that the EMTC method provides appreciable accuracy under various P-G fault conditions within a complex network topology. Our findings suggest that it is safe to use constant-parameter models in the simulation of pre-calculation, without compromising the accuracy of fault location using the EMTC method or EMTR-based approaches, such as [28].

## C. A network coinciding with IEEE 9-bus System

This case studies whether the method is valid for the cases that one line end is a busbar connected with another line segment instead of a transformer. A network with the same structure as IEEE 9-bus system, as shown in Fig. 16, is taken as an example. The line parameters and fault conditions are consistent with section IV B. Several P-G faults on line 9-6 are studied. The fault-generated transients are monitored on bus 9 and the GFLs are set on each line with a resolution of 10 m.

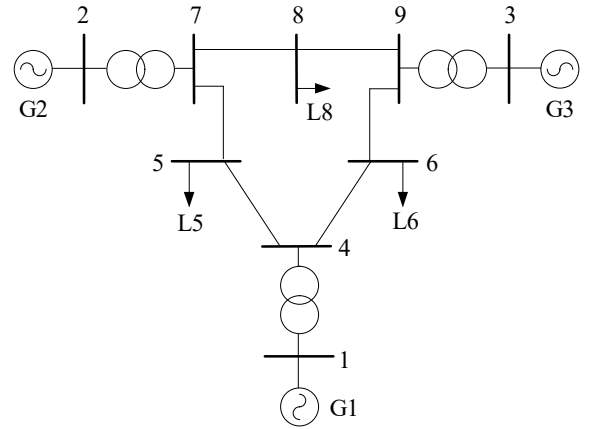

Fig. 16. IEEE 9-bus system. Line 9-6 is assumed 20 km in length.

TABLE. VIII. FAULT LOCATION RESULTS AND ERRORS IN IEEE 9-BUS SYSTEM

| Distance from bus 9/km | Located faulty line | Location result/km | Error/m |
|---|---|---|---|
| 2 | 9-6 | 1.99 | 10 |
| 5 | 9-6 | 4.89 | 110 |
| 8 | 9-6 | 7.96 | 40 |
| 10 | 9-6 | 9.93 | 70 |
| 13 | 9-6 | 12.95 | 50 |
| 16 | 9-6 | 15.87 | 130 |
| 19 | 9-6 | 18.87 | 130 |

The results shown in Table VIII demonstrate that EMTC identifies the right faulty line and the accuracy agrees with Table VII. In addition, it is not essential to place the measurements at the end of the faulty line. External transients (e.g., on bus 8) can also be adopted for fault location.

## V. Conclusion

In this study, we proposed a novel method for fault location in power networks utilizing electromagnetic transient convolution (EMTC), considering the influence of frequency-dependent power line parameters and lossy ground.

Through our evaluation of the EMTC method, we observed significant location errors when using a naive approach that employed frequency-dependent parameters in real fault simulation, but constant parameters in pre-calculation. The location errors exhibited a roughly linear growth pattern with increasing fault distance.

To overcome the issue in multi-phase line cases when using the naive EMTC approach, we employed a phase-mode transformation and performed EMTC in aerial modes. Through this approach, we were able to greatly reduce location errors since the frequency-dependence and ground loss could be neglected in aerial modes.

To evaluate the effectiveness of the aerial mode EMTC approach in reducing location errors in transmission lines with frequency-dependent parameters and lossy ground, we conducted numerical experiments in a 220-kV 300-km transmission line, a radial distribution network and IEEE 9-bus system. Our results demonstrate that the location errors can be greatly reduced using the aerial mode transient approach. These findings provide valuable insight into the potential use of this approach for improving fault location accuracy in diverse power network scenarios and conditions.